\documentclass[3p, twocolumn,sort&compress]{elsarticle}
\journal{Physica Medica}

\usepackage[english]{babel}
\usepackage{graphicx,tabularx,booktabs,colortbl}
\usepackage{amsfonts,amsmath,amssymb,bm}
\usepackage{isotope}
\usepackage[numbers]{natbib}

\usepackage{textcomp}
\usepackage{hyperref}

\newcommand{\gb}{\, \mathrm{GBq} }
\newcommand{\mb}{\, \mathrm{MBq} }
\newcommand{\kev}{\, \mathrm{keV} }

\begin{document}

%%% Start of article front matter
\begin{frontmatter}

\title{Phantom study for \isotope[90]{Y} liver radioembolization dosimetry with a long axial field-of-view PET/CT}

\author[1]{Lorenzo Mercolli\corref{cor1}}  \ead{lorenzo.mercolli@insel.ch}
\author[1]{Konstantinos Zeimpekis }  
\author[1]{George A. Prenosil }  
\author[2]{Hasan Sari} 
\author[1]{Hendrik G. Rathke}  
\author[1]{Axel Rominger}
\author[1]{Kunagyu Shi}  

\cortext[cor1]{Corresponding author}
\address[1]{Department of Nuclear Medicine, Inselspital \\ Bern University Hospital, University of Bern \\Freiburgstrasse 18, CH-3010 Bern \\ Switzerland }
\address[2]{Advanced Clinical Imaging Technology \\ 
Siemens Healthcare AG \\
Lausanne \\Switzerland
}

\begin{abstract} % abstract
  \textbf{Purpose:} The physical properties of yttrium-90 (\isotope[90]{Y}) allow for imaging with positron emission tomography/computed tomography (PET/CT). The increased sensitivity of long axial field-of-view (LAFOV) PET/CT scanners possibly allows to overcome the small branching ratio for positron production from \isotope[90]{Y} decays and to improve for the post-treatment dosimetry of \isotope[90]{Y} of selective internal radiation therapy.

  \textbf{Methods:} For the challenging case of an image quality body phantom, we compare a full Monte Carlo (MC) dose calculation with the results from the two commercial software packages Simplicit90Y and Hermes. The voxel dosimetry module of Hermes relies on the \isotope[90]{Y} images taken with a LAFOV PET/CT, while the MC and Simplicit90Y dose calculations are image independent.

  \textbf{Results:} The resulting doses from the MC calculation and Simplicit90Y agree well within the error margins. The image-based dose calculation with Hermes, however, consistently underestimates the dose. This is due to the mismatch of the activity distribution in the PET images and the size of the volume of interest. We found that only for the smallest phantom sphere there is a statistically significant dependence of the Hermes dose on the image reconstruction parameters and scan time.

  \textbf{Conclusion:} Our study shows that Simplicit90Y's local deposition model can provide a reliable dose estimate. On the other hand, the image based dose calculation suffers from the suboptimal reconstruction of the \isotope[90]{Y} distribution in small structures. 
\end{abstract}

  
\end{frontmatter}

%\section*{todo}

%\begin{itemize}
%    \item Cite \cite{costa2021} (explorer with Y90), check also \cite{mikell2018}
%    \item rewrite abstract
%\end{itemize}

%%%%%%%%%%%%%%%%%%%%%%%%%%%%%%%%%%%%%%%%%%%%%%%%%%%%%%%%%%%%%%%%%%%%%%%%%%%%%%%%
\section*{Background}

Nowadays, yttrium-90 (\isotope[90]{Y}) selective internal radiation therapy (SIRT) is a well-established and effective treatment modality for hepatocellular carcinoma (HCC) and liver metastases of neuroendocrine tumours as well as colorectal cancer \cite{levillain2021,vilgrain2017}. SIRT exploits the fact that the blood supply of hepatic malignancies is different compared to normal liver parenchyma. % \cite{Ahmadzadehfar2010}. 
It is therefore possible to target the tumor cells locally through the injection of \isotope[90]{Y} microspheres into the arteries of the liver.

Individual dosimetry and treatment planning play a pivotal role in applying SIRT safely and for achieving the best possible treatment outcome. An individual treatment planing and verification is recommended by the EANM position paper \cite{Konijnenberg2021}. For the pre-treatment dosimetry of SIRT, different methods are part of the clinical standard procedure \cite{Dezarn2011,Giammarile2011,tafti2019,dieudonne2016}. However, despite being recommended \cite{Dezarn2011,Giammarile2011}, there is not yet a standardized protocol for post-treatment verification in the clinical routine \cite{Brosch2020,tafti2019}.

The known challenges of imaging \isotope[90]{Y} Bremsstrahlung \cite{Elschot2013,tafti2019,Dezarn2011} with single photon emission computed tomography (SPECT) made the use of positron emission tomography (PET) systems an increasingly popular alternative to SPECT \cite{Lhommel2009} and is now the recommended procedure for resin microspheres \cite{levillain2021}. Post-treatment verification with PET exploits the fact that \isotope[90]{Y} can decay to the excited $O^+$ state of \isotope[90]{Zr} \cite{nds2020}. This state can then further decay through an internal pair production from an $E0$ transition with a branching ratio (BR) of $(3.26 \pm 0.04)\cdot 10^{-6}$ $e^+ e^-$ pairs per decay \cite{Dryak2020}. Despite this low BR, Ref.~\cite{Lhommel2009} showed more than a decade ago that a post-treatment verification with PET is feasible.

Recently, long axial field-of-view (LAFOV) PET/CT systems have found their way into clinical routine \cite{Cherry2018,Badawi2019, Spencer2021,prenosil2022,dai2023, Alberts2021}. The sensitivity and noise equivalent count rate of LAFOV PET/CT systems significantly improves over standard field-of-view systems. For the imaging of \isotope[90]{Y} microspheres, LAFOV PET/CT outperform conventional PET/CT systems since it can compensate the low branching fraction of the excited \isotope[90]{Zr} state \cite{zeimpekis2023}. Furthermore, the limited FOV of conventional PET/CT systems and the resulting decrease in sensitivity towards edges of the FOV makes the imaging of the whole liver with a single bed position very challenging.

The aim of this study is to compare a dose calculation method that is based on LAFOV PET images with image independent methods. To this end, we performed phantom scans and computed the dose using two commercial software products: Hermes (Hermes Medical solutions, Stockholm, Sweden) and Simplicit90Y (Mirada Medical Ltd, Oxford, UK; Boston Scientific Corporation, Marlborough, MA, USA). Hermes' voxel dosimetry module uses a so-called semi-Monte Carlo (sMC) algorithm for the dose calculation and therefore requires quantitative PET images as input. Simplicit90Y computes the deposited dose assuming a local deposition model, which for the case of a phantom is independent of the PET image. Finally, we performed a Monte Carlo (MC) simulation with the software FLUKA \cite{fluka1,fluka2,flair}, which provides the full modelling of the \isotope[90]{Y} decay, particle transport and dose deposition. It is completely independent of the imaging and segmentation process and therefore serves as the ground truth for the dose calculation.

%%%%%%%%%%%%%%%%%%%%%%%%%%%%%%%%%%%%%%%%%%%%%%%%%%%%%%%%%%%%%%%%%%%%%%%%%%%%%%%%
\section*{Materials and methods}

\subsection*{Measurements and imaging protocols}

For the image acquisition of \isotope[90]{Y} with a LAFOV PET/CT, two NEMA International Electrotechnical Commission (IEC) body phantoms (Data Spectrum Corp.) \cite{phantom,nema} were filled with demineralized water, a minor addition of hydrochloric acid and \isotope[90]{Y} citrate. The total activity at reference time was $1.31 \pm 0.20\gb$, while at scan time it had decayed to $1.12 \pm 0.17\gb$. All activity measurements were carried out with an ISOMED 2010 well-type dose calibrator that was calibrated by the Swiss Federal Institute of Metrology METAS. The activity measurement of almost pure $\beta$ emitters like \isotope[90]{Y} is challenging, in particular with well-type dose calibrators. In order to take into account the systematic error of the activity measurements (calibration factor, variations in the measurement geometry, filling levels, etc.), we assume an error of $15\%$, which is within the maximum admissible error for a calibrated system in Switzerland (see e.g.\ Ref.~\cite{bochud2006dose}). 

One phantom had a hot and the second one a cold background. In Tab.~\ref{t:A_phantom} we report the activities inside the volumes of the six spheres and the background volumes of the phantoms. The sphere to background activity concentration ratio is approximately $1:10$ with an activity concentration of $\approx 1.3 \, \mb / \mathrm{ml}$ in the spheres. The activities in the phantoms' spheres were lower compared to typical lesions that are treated with SIRT. The setup should therefore be considered as a rather challenging case for imaging and dose calculations. 

\begin{table*}[htb]
\centering
  \begin{tabular}{lccc}
    \toprule
     Volume name & Diameter $[\mathrm{mm}]$ & Volume $[\mathrm{ml}]$ & Activity $[\mathrm{MBq}]$ \\
     \midrule
    cold bkg & N/A & $9762.0 \pm 98.5 $ & $ 0.0$ \\
    hot bkg & N/A & $9829.0 \pm 97.8$ & $ 1193.51 \pm 0.18$ \\
    $s_1$ & $10.0 \pm 0.5$ & $0.528 \pm 0.079 $ & $0.64 \pm 0.15$ \\
     $s_2$ & $13.0 \pm 0.5$ &$1.16 \pm 0.13$ & $1.40 \pm 0.27$ \\
    $s_3$ & $17.0 \pm 0.5$ &$2.58 \pm 0.23$ & $3.12 \pm 0.55$ \\
     $s_4$ & $22.0 \pm 1.0$ &$5.61 \pm 0.76$ & $6.78 \pm 1.40$\\ 
    $s_5$ & $28.0 \pm 1.0$ & $11.5 \pm 1.2$ & $13.9 \pm 2.60$ \\
     $s_6$ & $37.0 \pm 1.0$ & $26.6 \pm 2.2$ & $ 32.1 \pm 5.60$ \\
    \bottomrule
  \end{tabular}
  \caption{\isotope[90]{Y} activities in at reference time and the dimensions of the background and sphere volumes of the NEMA IQ phantoms. The quoted errors are described in the text.  }\label{t:A_phantom}
\end{table*}

The error on the sphere volumes are derived from the errors on the diameters quoted in the NEMA phantom specifications. These are the possible variations in the production process of the phantoms. The background volumes are determined through their weight when filled with demineralized water and we assume a $1\%$ error on the weight measurement and a water density of $0.9982 \, \mathrm{g}/\mathrm{ml}$ according to the ICRU Report 90 \cite{icru90}.

The two phantoms were imaged with a Biograph Vision Quadra PET/CT scanner (Siemens Healthineers, Knoxville, TN, USA) at the Department of Nuclear Medicine, Inselspital, Bern University Hospital. The detectors of the Biograph Vision Quadra are made of $5\times 5$ arrays of $3.2 \times 3.2 \times 20 \, \mathrm{mm}$ lutetium oxyorthosilicate (LSO) crystals that are coupled to a $16\times 16 \, \mathrm{mm}$ array of silicon photomultipliers. $4 \times 2 $ such mini-blocks are arranged into one detector block. The whole PET detector consists of 32 detector rings with 38 blocks each. This yields an axial FOV of $106 \, \mathrm{cm}$. All images were acquired with a maximum ring distance of 322 crystals (MRD 322) and reconstructed the images in ultra-high sensitivity mode (UHS). For comparison, images were also reconstructed in high sensitivity mode (HS) with a MRD of 85 crystals. The NEMA sensitivity for \isotope[18]{F} in UHS is $176 \,\mathrm{kcps}/\mb$, while in HS it is $ 83 \, \mathrm{kcps}/\mb$ \cite{prenosil2022}. The time-of-flight (TOF) resolution of the Biograph Vision Quadra is $228$ and $230$ for HS and UHS, respectively. While the Biograph Vision Quadra's sensitivity exceeds the one of standard FOV scanners, the TOF and spatial resolution of $3.3 \times 3.4 \times 3.8 \, \mathrm{mm}$ are comparable to standard systems. 

\begin{figure*}[htb]
  \centering 
  \includegraphics[width=0.4\textwidth]{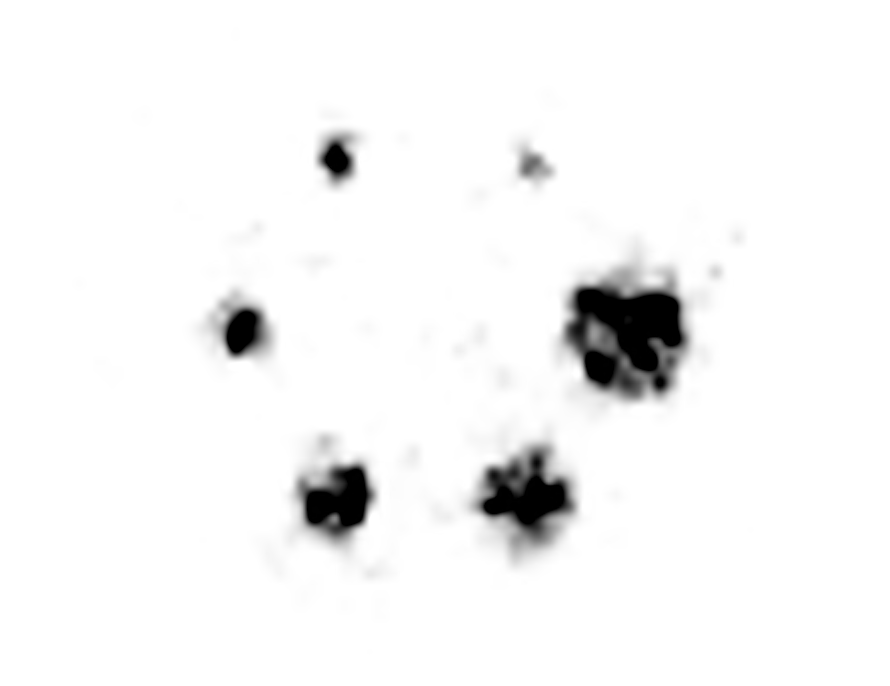}
  \hfill 
  \includegraphics[width=0.4\textwidth]{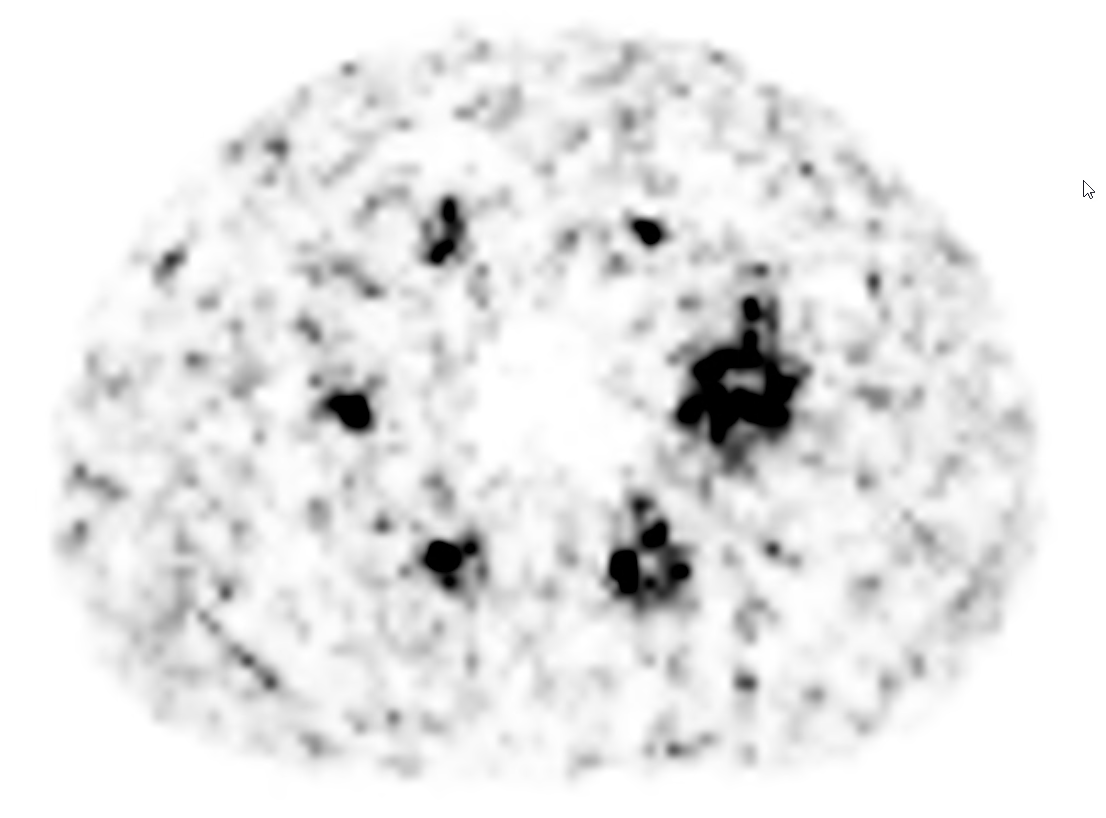} \\
  \includegraphics[width=0.4\textwidth]{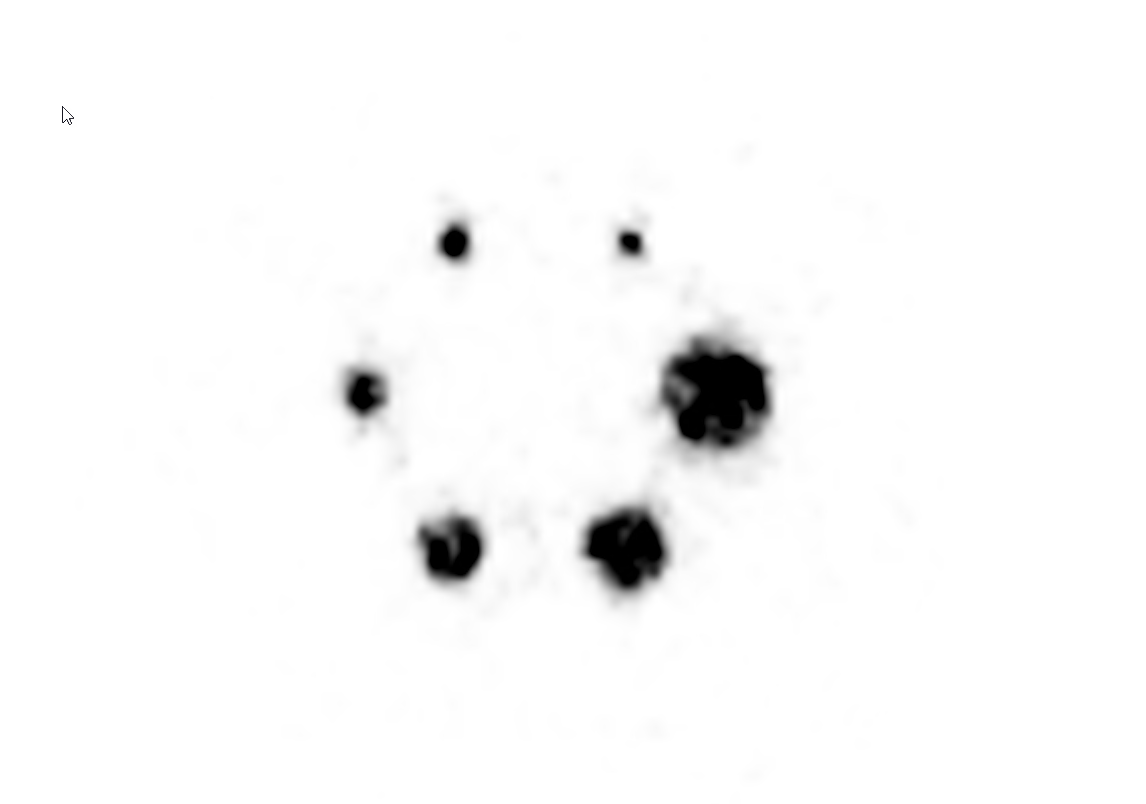}
  \hfill 
  \includegraphics[width=0.4\textwidth]{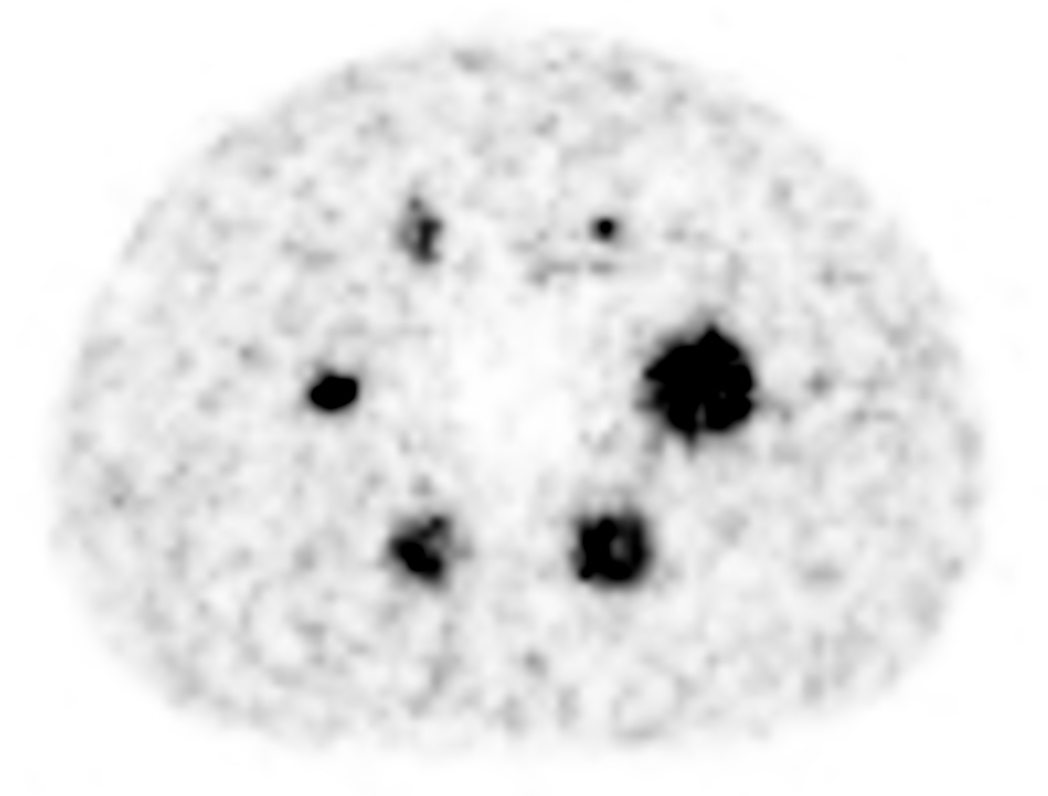}
  \caption{Example images of UHS reconstruction and with $5 \, \mathrm{min}$ (top) and $20 \, \mathrm{min}$ (bottom) acquisition time for both cold and hot backgrounds. }\label{f:uhs_5_20min}
\end{figure*}

The images of the two phantoms were acquired during $50 \,\mathrm{min}$. Reconstructions were performed using a dedicated image reconstruction prototype (e7-tools, Siemens Healthineers). The error on the resulting dose does not include any error form the image reconstruction process. 
For the image reconstruction Siemens' TOF, point-spread function (PSF) recovery and ordered subset expectation maximization (OSEM) is used. According to Ref.~\cite{zeimpekis2023}, the optimal reconstruction parameters in terms of image quality for both HS and UHS are two iterations, five subsets, a $2\, \mathrm{mm}$ Gaussian filter and a $220\times 220$ matrix. In order to assess the dependence dose calculation's dependence on the scan time, images were reconstructed with the first $5$, $10$ and $20 \, \mathrm{minutes}$ of scan duration in UHS by rebinning the list-mode data. For comparison, we also reconstructed an image in HS mode with $20\, \mathrm{minutes}$ scan duration. Fig.~\ref{f:uhs_5_20min} shows examples with UHS reconstruction and with $5 \, \mathrm{min}$ and $20\, \mathrm{min}$ acquisition time. 
The decay correction and quantification for \isotope[90]{Y} is done directly by the vendor's image acquisition and reconstruction software. Of course, an uncertainty in the quantification of the activity concentration in the PET image could propagate to the image based dose calculation. Based on the recovery coefficients in Ref.~\cite{zeimpekis2023} for the larger spheres and the reconstruction protocol described above, we assume a $10\%$ error on the PET image quantification. Note that this is not included the $15\%$ error from the activity measurement.

\subsection*{Dose calculations}

The doses deposited in the phantom's spheres were calculated using three independent methods: a full MC simulation of the NEMA phantom, the image based sMC method from Hermes' voxel dosimetry module and Simplicit90Y's local deposition model. %In SIRT, and in particular in phantom studies, it is safe to assume that the microspheres do not move and that therefore one can simply consider the physical decay constant of \isotope[90]{Y} in order to compute the total number of decays inside the phantom spheres. We did not consider the error of the \isotope[90]{Y} decay constant in our analysis since it is $0.08 \%$ according to Ref.~\cite{nds2020}. 

\begin{figure*}[htb]
  \centering 
  \includegraphics[width=0.45\textwidth]{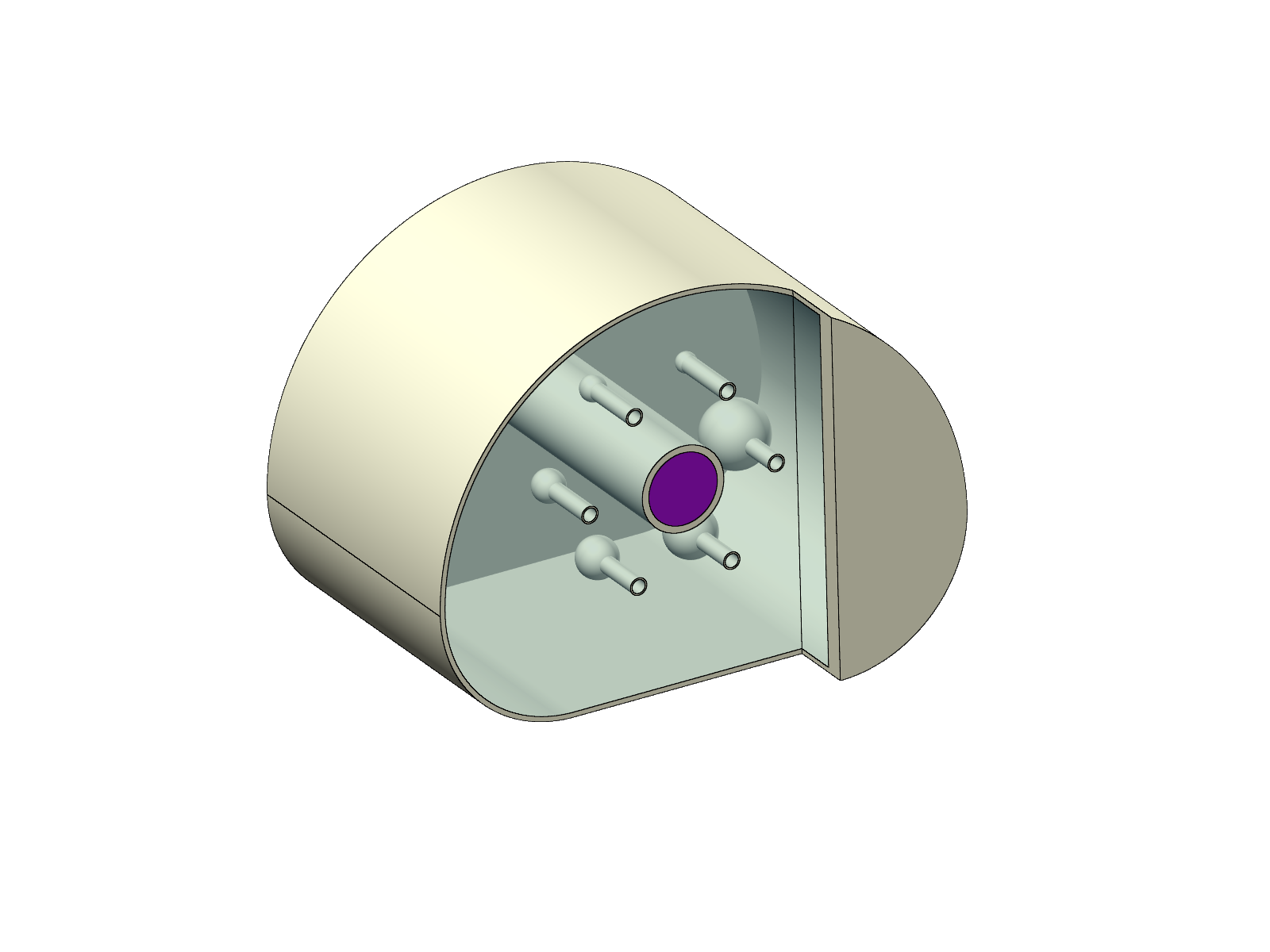}
  \hfill 
  \includegraphics[width=0.45\textwidth]{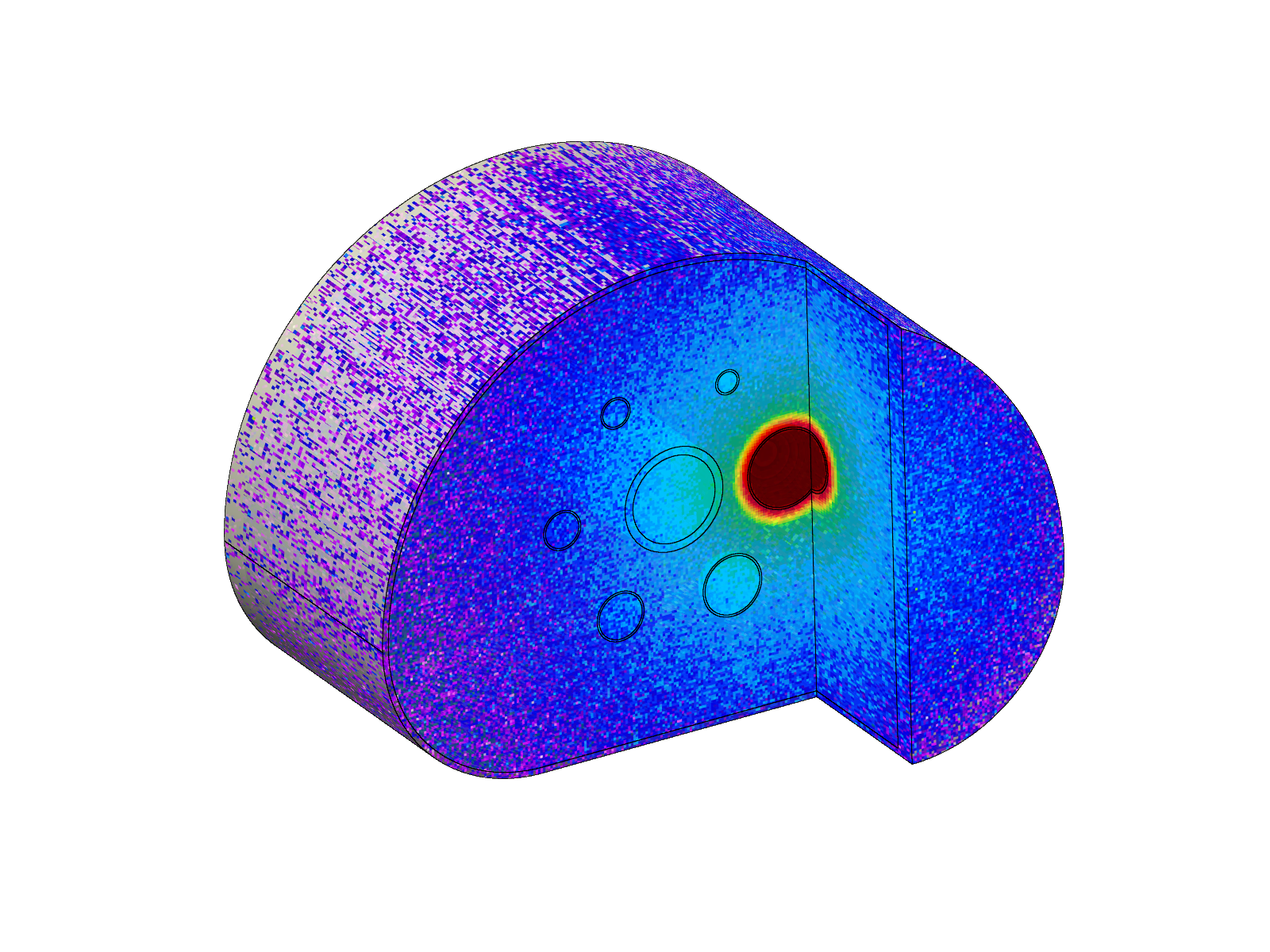}
  \caption{FLUKA implementation of the NEMA IQ phantom (left) and an illustrative example of a 3D dose distribution from FLUKA with cartesian grid scoring (right). }\label{f:fluka_geo}
\end{figure*}

% MC
The NEMA phantom was implemented in CERN's FLUKA 4-2.2 \cite{fluka1,fluka2} and Flair 3.1-15.1 \cite{flair}, which is a general MC framework for particle transport. Fig.~\ref{f:fluka_geo} shows the rendering of the phantom geometry and an example of the dose distribution. %The simulated phantom is filled with water from FLUKA's built-in material database. 
We used the standard \mbox{PRECISIO} setting of FLUKA for the physics and transport parameters (most importantly, the particle transport threshold is set to $100 \, \kev$). The simulations were run in semi-analogue mode, i.e.\ the \isotope[90]{Y} decay is simulated with random decay times, daughters and inclusive decay radiation, for every single sphere and the hot background separately. The simulation of the hot background required a dedicated source routine due to the non-standard shape of the background volume. The decay properties of \isotope[90]{Y} are taken in FLUKA's isotope library. % and there is no need for any manual specification of particle spectra. 
The resulting doses were scored with the USRBIN routine in spherical regions (region binning) with the nominal size NEMA phantom spheres. This avoids volume errors from voxelization. Of course, the simulations's results need to be normalized to the measured total activities in Tab.~\ref{t:A_phantom}. The simulation of $10^7$ primaries were sufficient for reaching a negligible statistical error in the regional dose scoring and we do not assume any systematic error in the simulation itself (geometry implementation, \isotope[90]{Y} decay data, transport thresholds, etc.). Therefore, the uncertainties on the doses from the MC simulations stem only from the uncertainties related to the activity measurements shown in Tab.~\ref{t:A_phantom}. 

% Hermes
Hermes' voxel dosimetry module provides the possibility to calculate a dose map from a quantitative PET or SPECT image through the so called sMC method \cite{hippelaeinen2015,hippelaeinen2017}. % \cite{heikkonen2016,hippelaeinen2016}
The authors of Ref.~\cite{hippelaeinen2017} proposed the sMC for the post-treatment dosimetry of lutetium-177 (\isotope[177]{Lu}) peptide receptor radionuclide therapy in order to overcome the accurate but time consuming full MC calculations (see also Refs.~\cite{hippelaeinen2015,hippelaeinen2016}). The basic idea behind sMC is to separate the electron and photon transport: the electron's energy is absorbed locally, i.e.\ in the same voxel where the decay has been detected, while for the photons a point-wise transport is used. The algorithm first converts the activity distribution from and image to an electron dose map. Second, the photons are transported in a full MC down to a threshold of $15 \kev$ \cite{hippelaeinen2015}. The energy of photons with an energy below the transport threshold as well as the energy of recoil electrons is deposited locally. Note that for the photon transport sMC uses only cross sections of water, which are then rescaled to the various tissue densities \cite{hippelaeinen2017}. %This might lead to a significant limitation when computing doses in bones, since scaling with the atomic number is not taken into account \cite{hippelaeinen2015}. However, for SIRT this is only marginally relevant. 

% This ingenious method has been applied and successfully validated for peptide receptor radionuclide therapies with \isotope[177]{Lu} \cite{hippelaeinen2015,hippelaeinen2016,hippelaeinen2017}.
With a mean electron energy of $\bar{E}_\beta = 148.8 \kev$ ($79.44 \%$ intensity) and $Q_\beta = 496.8 \kev$ in conjunction with the low spatial resolution of SPECT images, \isotope[177]{Lu} is an ideal radionuclide for sMC. However, the literature on sMC for other radionuclides, such as \isotope[90]{Y} or \isotope[131]{I}, is scarce and, as pointed out the authors of Ref.~\cite{hippelaeinen2015}, further investigation for the applicability of sMC beyond \isotope[177]{Lu} is necessary. 

The sMC implementation of Hermes allows for a truly image based voxel-wise dose calculation. Therefore, different scan times and reconstruction settings can lead to different doses. For every input image (in units of $\mathrm{Bq}/\mathrm{ml}$) Hermes computes a dose map, i.e.~an image with units of $\mathrm{Gy}$. For all dose maps, we used Hermes Hybrid Viewer to segment the phantom spheres. The volumes-of-interest (VOI) were drawn on the CT according to to the nominal sphere diameters of the NEMA phantom. Since the dose scales with the inverse volume, segmenting the phantom spheres in the dose maps by a threshold (e.g. relative to some maximal voxel value) would make a comparison with the other dose calculation methods meaningless.  

Hermes does not provide information on the statistical error of the voxel values of the dose map from the sMC method. As mentioned before, we do not consider any error from the image reconstruction but add a $10\%$ uncertainty on the quantification. %Hence, the error of the dose calculation with Hermes we quote in Tabs.~\ref{t:doses_cold} and \ref{t:doses_cold} stems only from the voxelization of the sphere volumes.
Hermes' Hybrid Viewer reports the volume of each (voxelized) VOI and the deviation from the nominal spheres can reach $10\%$. 

In order to disentangle the systematic error from the sMC algorithm from the uncertainty of the activity distribution in the PET image, we constructed a synthetic PET image and passed it through Herme's voxel dosimetry module. This synthetic PET image contains the spheres of the NEMA phantom with the the nominal activities from Tab.~\ref{t:A_phantom} as well as the hot background. In the systematic error of the resulting doses, only the voxelization of the spheres stems from the synthetic PET image while the rest can be attributed to the sMC algorithm.

% Simplicit90Y
Lastly, the dose inside the phantom spheres was also computed with Simplicit90Y's multi-compartment dosimetry module (see e.g.\ Refs.~\cite{lotter2021,dieudonne2016,Dezarn2011} for multi-compartment dosimetry). Simplicit90Y relies on a local deposition model with a homogeneous activity distribution \cite{simplicitymanual}. The dose in a compartment or perfused volume is computed according to  

\begin{equation}
  \bar{D}_c \; = \; 50 \cdot  \frac{A_c \cdot (1-F) \cdot (1-R)}{m} \;, 
\end{equation}
where $A_c$ is the activity in $\mathrm{GBq}$ inside the compartment and $m$ is its mass in $\mathrm{kg}$. In the phantom measurements the lung shunt fraction $F$ and the residual waste fraction $R$ were set to zero. According to the Simplicit90Y manual the dose factor is fixed at $50 \, \mathrm{Gy} \, \mathrm{kg} / \mathrm{GBq}$. Note that slightly lower values can be found in the literature (see e.g.\ Refs.~\cite{dieudonne2016,Dezarn2011}). In Simplicit90Y, all dose calculations assume a liver density of $1.06 \, \mathrm{g}/\mathrm{cm}^3$. Since in the phantoms are filled with water, we corrected the doses from Simplicit90Y by the ratio of the liver and water density. Simplicit90Y only takes image information if there are overlapping volumes. This is not the case for the NEMA phantoms and therefore the dose calculation with Simplicit90Y is independent of the PET image. The constant dose factor is simply multiplied by the activities in Tab.~\ref{t:A_phantom}. As in the case of the sMC dose calculation, the VOI were drawn as spheres with the size according to the nominal sphere diameters. There is some deviation from the exact spherical volume due to the voxelization of the volume (in the CT image) and possibly some rounding errors of Simplicit90Y. The uncertainties of the dose values from Simplicit90Y are therefore composed of a $10\%$ error from the volume voxelization and the error on the activities inside the spheres in Tab.~\ref{t:A_phantom}. We do not consider any systematic uncertainty on the local deposition model nor on the constant dose factor of $50 \, \mathrm{Gy} \, \mathrm{kg} / \mathrm{GBq}$.

%%%%%%%%%%%%%%%%%%%%%%%%%%%%%%%%%%%%%%%%%%%%%%%%%%%%%%%%%%%%%%%%%%%%%%%%%%%%%%%%
\section*{Results} 

\begin{figure*}[htb]
  \centering 
  \includegraphics[width=0.75\textwidth]{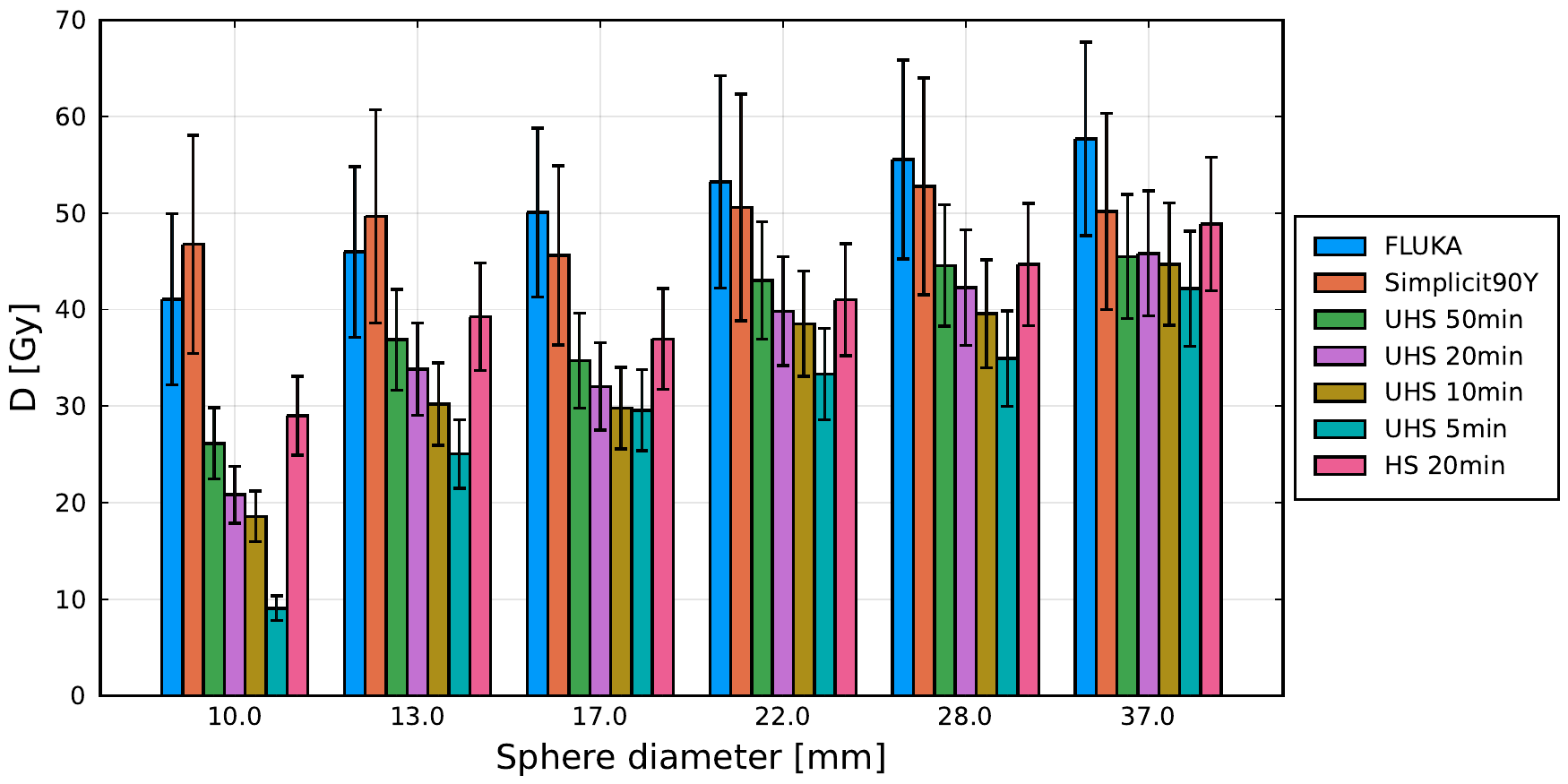} \\[2ex]
    \includegraphics[width=0.75\textwidth]{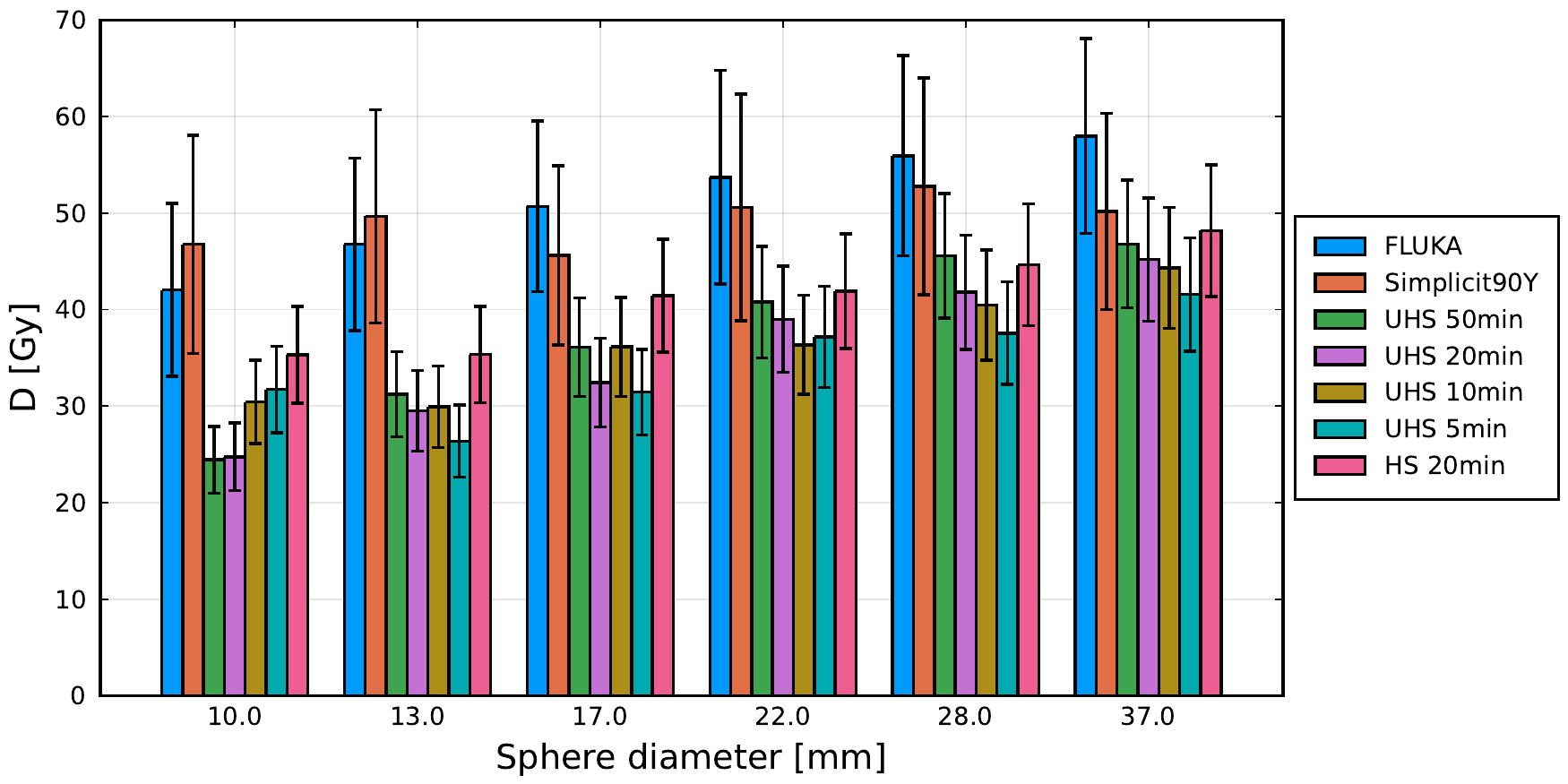}
  \caption{Visualization of the doses in Tabs.~\ref{t:doses_cold} and \ref{t:doses_hot} for the cold (top) and cold (bottom) background phantoms.}\label{f:hist_doses}
\end{figure*}

Tabs.~\ref{t:doses_cold} and \ref{t:doses_hot} report the doses deposited in the six spheres of the cold and hot background phantoms. Fig.~\ref{f:hist_doses} visualizes the values in Tabs.~\ref{t:doses_cold} and \ref{t:doses_hot}. As previously mentioned, the FLUKA dose calculation is considered as the ground truth for the doses.

\begin{table*}[htb]
\centering
  \begin{tabular}{lllllll}
    \toprule
    Method   & $D_{s_1} \, [\mathrm{Gy}]$   & $D_{s_2} \, [\mathrm{Gy}]$    & $D_{s_3} \, [\mathrm{Gy}]$  & $D_{s_4} \, [\mathrm{Gy}]$  & $D_{s_5} \, [\mathrm{Gy}]$    & $D_{s_6} \, [\mathrm{Gy}]$  \\
    \midrule 
  FLUKA      & 41.1 ± 8.9  & 46.0 ± 8.8  & 50.1 ± 8.8 & 53.2 ± 11.0 & 55.6 ± 10.0 & 57.7 ± 10.0 \\
Simplicit90Y & 46.8 ± 11.0 & 49.7 ± 11.0 & 45.6 ± 9.3 & 50.6 ± 12.0 & 52.8 ± 11.0 & 50.2 ± 10.0 \\
UHS 50 min    & 26.2 ± 3.7  & 36.9 ± 5.2  & 34.7 ± 4.9 & 43.0 ± 6.1  & 44.6 ± 6.3  & 45.5 ± 6.4  \\
UHS 20 min     & 20.8 ± 2.9  & 33.8 ± 4.8  & 32.0 ± 4.5 & 39.8 ± 5.6  & 42.3 ± 6.0  & 45.8 ± 6.5  \\
UHS 10 min     & 18.6 ± 2.6  & 30.2 ± 4.3  & 29.8 ± 4.2 & 38.5 ± 5.4  & 39.6 ± 5.6  & 44.7 ± 6.3  \\
UHS 5 min      & 9.07 ± 1.3  & 25.1 ± 3.5  & 29.6 ± 4.2 & 33.3 ± 4.7  & 35.0 ± 4.9  & 42.2 ± 6.0  \\
HS 20 min      & 29.0 ± 4.1  & 39.3 ± 5.6  & 37.0 ± 5.2 & 41.0 ± 5.8  & 44.7 ± 6.3  & 48.9 ± 6.9 \\
    \bottomrule
    \end{tabular}
  \caption{Doses in $\mathrm{Gy}$ in the six spheres of the NEMA phantom with a cold background according to the different dose calculation methods.}\label{t:doses_cold}
\end{table*}

\begin{table*}[htb]
\centering 
  \begin{tabular}{lllllll}
    \toprule
    Method   & $D_{s_1} \, [\mathrm{Gy}]$   & $D_{s_2} \, [\mathrm{Gy}]$    & $D_{s_3} \, [\mathrm{Gy}]$  & $D_{s_4} \, [\mathrm{Gy}]$  & $D_{s_5} \, [\mathrm{Gy}]$    & $D_{s_6} \, [\mathrm{Gy}]$  \\
    \midrule
   FLUKA      & 42.0 ± 9.0  & 46.8 ± 8.9  & 50.7 ± 8.8 & 53.7 ± 11.0 & 55.9 ± 10.0 & 58.0 ± 10.0 \\
    Simplicit90Y & 46.8 ± 11.0 & 49.7 ± 11.0 & 45.6 ± 9.3 & 50.6 ± 12.0 & 52.8 ± 11.0 & 50.2 ± 10.0 \\
    UHS 50 min     & 24.5 ± 3.5  & 31.2 ± 4.4  & 36.1 ± 5.1 & 40.8 ± 5.8  & 45.6 ± 6.4  & 46.8 ± 6.6  \\
UHS 20 min     & 24.8 ± 3.5  & 29.5 ± 4.2  & 32.5 ± 4.6 & 39.0 ± 5.5  & 41.8 ± 5.9  & 45.2 ± 6.4  \\
UHS 10 min     & 30.4 ± 4.3  & 30.0 ± 4.2  & 36.1 ± 5.1 & 36.4 ± 5.1  & 40.5 ± 5.7  & 44.3 ± 6.3  \\
UHS 5 min      & 31.7 ± 4.5  & 26.4 ± 3.7  & 31.4 ± 4.4 & 37.2 ± 5.3  & 37.6 ± 5.3  & 41.6 ± 5.9  \\
HS 20 min      & 35.3 ± 5.0  & 35.4 ± 5.0  & 41.4 ± 5.9 & 41.9 ± 5.9  & 44.6 ± 6.3  & 48.2 ± 6.8  \\
  \bottomrule
  \end{tabular}
  \caption{Doses in $\mathrm{Gy}$ in the six spheres of the NEMA phantom with a hot background according to the different dose calculation methods.}\label{t:doses_hot}
\end{table*}

Fig.~\ref{f:hist_doses} shows an increase of dose from the MC simulation with increasing sphere diameter. In the case of the hot phantom this effect is slightly less pronounced. However, the error bars are large, due to the uncertainty of the activity measurement, and therefore this effect is statistically insignificant. The Simplict90Y doses vary depending on the sphere diameters and are exactly the same for both phantoms. 

Compared to FLUKA and Simplicit90Y, the Hermes based dose calculations shows a systematic underestimation of the doses. This underestimation is particularly pronounced in the case of the two smallest spheres and for short acquisition times. In most cases the dose from the UHS images decreases with decreasing acquisition time. Interestingly, the HS based dose is comparable or even closer to the FLUKA dose than the dose from the and 50 min UHS image. While most of the Hermes dose values lie still within the $1\sigma$ range of the FLUKA doses, the systematic uncertainties of the Hermes calculation require some discussion.

\begin{figure*}[htb]
    \centering
    \includegraphics[width=0.7\textwidth]{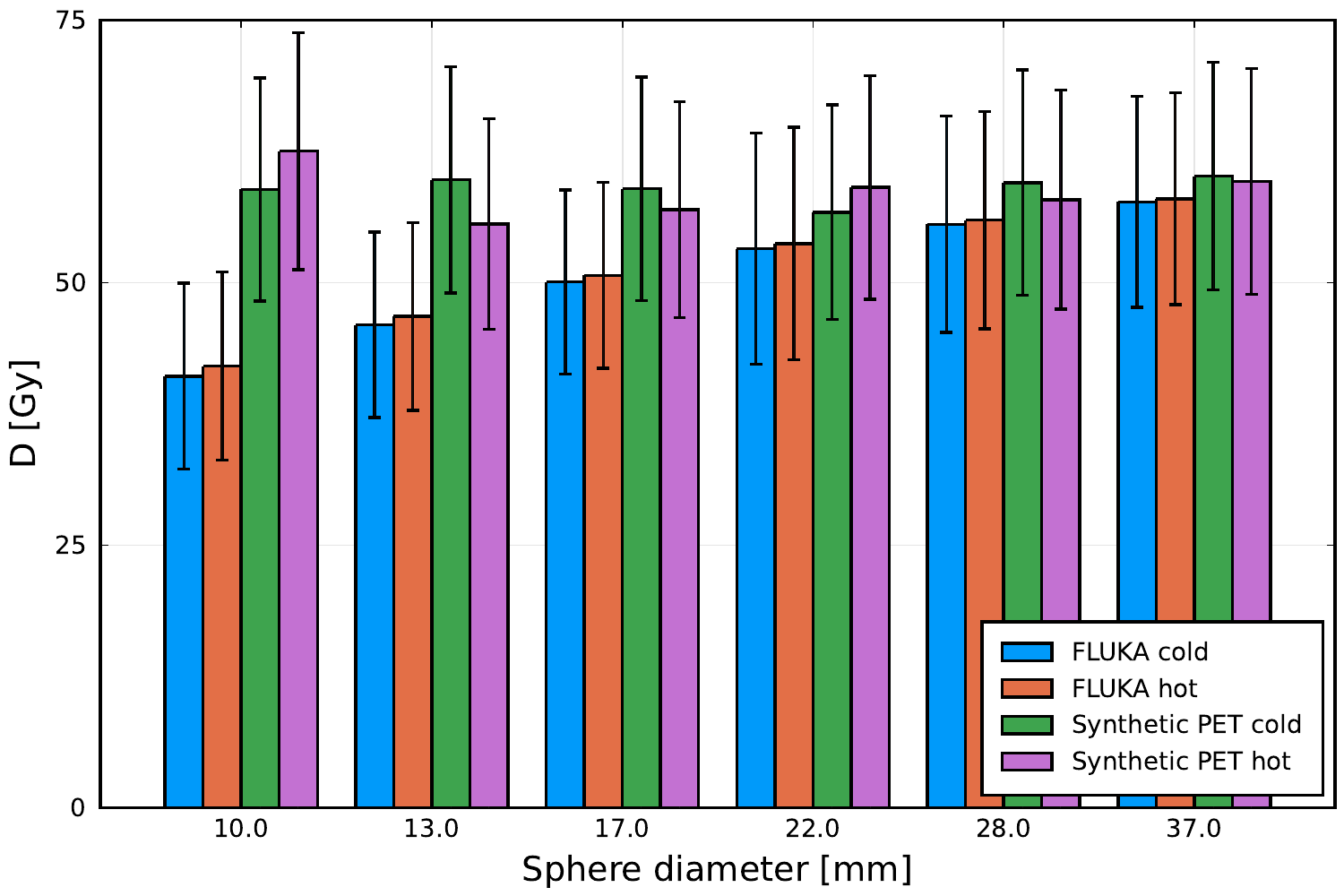}
    \caption{Comparison between the doses from FLUKA and the Hermes doses from the synthetic PET image.}\label{f:synthetic}
\end{figure*}

In Fig.~\ref{f:synthetic} we report the doses that stem from using a synthetic PET image with an optimal activity distribution as an input for Hermes' sMC algorithm. Compared to FLUKA, Hermes overestimates the doses for small sphere diameters.

%%%%%%%%%%%%%%%%%%%%%%%%%%%%%%%%%%%%%%%%%%%%%%%%%%%%%%%%%%%%%%%%%%%%%%%%%%%%%%%%
\section*{Discussion}

The increase of dose values from the MC simulation with increasing sphere diameter in Tabs.~\ref{t:doses_cold} and \ref{t:doses_hot} is due to the width of the \isotope[90]{Y} dose point kernel (DPK), i.e. the smaller the sphere the higher the relative fraction of the dose is deposited outside the sphere. When comparing FLUKA's doses of the cold and hot background phantoms, we can see a similar effect in the sense that with increasing sphere diameter the relative amount of the spill-in dose from the hot background becomes smaller. Note however, that the uncertainty due to the activity measurement is rather large, making these effects statistically insignificant. 

Since the \isotope[90]{Y} activity increases with the sphere volume but the dose decreases with the inverse volume, the doses from Simplicit90Y should in principle have the same value for all six spheres. However, due to the voxelization of the volumes in the CT images and the rounding errors of the volume in Simplicit90Y there is a variation in the doses for different sphere diameters in Fig.~\ref{f:hist_doses}. Nevertheless, the doses computed with Simplicit90Y are comparable the the FLUKA doses, in particular when taking into account the associated errors. Obviously, there is no difference between the cold and hot background doses since no information is taken from the PET images in Simplicit90Y in the absence of multiple compartments. In sum, the doses computed with Simplicit90Y confirm the fixed dose factor of $50 \, \mathrm{Gy} \, \mathrm{kg}/\mathrm{GBq}$ for larger volumes, hint towards issues with rounding errors for small volumes and show the inability to capture physical effects (width of \isotope[90]{Y} DPK) of the the dose distribution in small volumes.  

The doses values from Hermes in Fig.~\ref{f:hist_doses} offer an interesting comparison since they depend on the quantified PET images. Comparing the different scan times in UHS mode, Fig.~\ref{f:hist_doses} seems to show that the Hermes doses become smaller for shorter scan times. This is particularly pronounced for the case of the smallest sphere and the cold phantom. It is tempting to conclude that longer scan times lead to dose values closer to the ground trough. However, our analysis shows that this trend is not statistically significant due to the relatively large error bars. Even if the image quality improves with scan time in general, as shown in Ref.~\cite{zeimpekis2023}, the volume of the dose distribution remains small compared to the nominal sphere volume. Also Ref.~\cite{hesse2021} argues that longer scan times to not improve the results in such challenging imaging situations due to the LSO background radiation and the inherent blurring of \isotope[90]{Y} (imaged positrons stem from pair production of a high-energy prompt $\gamma$ in the \isotope[90]{Zr}). %We should also keep in mind that, as in the case of Simplicit90Y, the voxelization of the sphere volumes can lead to some variation in the dose values. % One could imagine to increase the matrix size of the PET image, but this would lead to a deterioration of image quality (see Ref.~\cite{zeimpekis2023}). 
Interestingly, the 20 minutes HS image seems to lead to doses that are comparable to the 50 minutes UHS image. %This supports further our conclusion that scan time and image quality play a minor role for the dose calculation. 

A priori, the systematic underestimation of the dose in the Hermes dose values compared to FLUKA can be either due to the quantification uncertainty in the PET image or due to the shortcomings of the sMC algorithm or both. Already the original authors pointed out that the assumptions underlying the sMC algorithm might not be applicable to \isotope[90]{Y} \cite{hippelaeinen2017}. However, the result from the synthetic PET image in Fig.~\ref{f:synthetic} shows that only for the smaller spheres, the sMC overestimates the deposited dose. Since the sMC algorithm deposits the $\beta^-$ dose in the source voxel, there high $\beta^-$ energy of \isotope[90]{Y} and the resulting dose spill-out, i.e.\ the width of the DPK, is not fully modelled. For the larger spheres, the dose spill-out becomes less relevant and the sMC results are consistent with FLUKA. This lets us conclude that Hermes' underestimation of the doses in Fig.~\ref{f:hist_doses} is dominated by the poor quantification of the activity distribution in the PET images. Indeed, the recovery coefficients in Ref.~\cite{zeimpekis2023} do not reach $100\%$, even for the largest sphere. 

% The authors of Ref.~\cite{costa2021} compared the full MC doses in a phantom that result from a voxelized synthetic PET image with an image obtained from a LAFOV PET system. Thei mask the 
Our results from the synthetic PET are in line with the results of Ref.~\cite{costa2021}. For smaller spheres the activity distribution is not quantified well enough, even with a LAFOV PET system. While in Ref.~\cite{costa2021} the a full MC dose calculation gives good results except for the smallest sphere, the sMC algorithm is limited to larger structures as shown in Fig.~\ref{f:synthetic}.

%%%%%%%%%%%%%%%%%%%%%%%%%%%%%%%%%%%%%%%%%%%%%%%%%%%%%%%%%%%%%%%%%%%%%%%%%%%%%%%%
\section*{Conclusions}

In this paper we compared different dose calculation methods for \isotope[90]{Y} radioembolization. While the full MC and Simplicit90Y dose calculations are independent of the input images, Hermes relies on the sMC algorithm to compute dose maps based on the activity distributions of images that were aquired with a LAFOV PET/CT. 

The MC and Simplicit90Y dose calculations agree well within the error margins. Simplicit90Y's local deposition model is not able to catch the effect of the width of the \isotope[90]{Y} DPK, which is most visible in small volumes. Furthermore, the voxelization of the exact sphere volume in Simplicit90Y leads to a nonphysical variation of the dose values for different sphere diameters.  

The Hermes' sMC method shows an underestimation of the dose for all input images compared to the full MC calculation. Processing a synthetic PET image revealed that while the sMC algorithm is unable to catch the width of the \isotope[90]{Y} DPK, the discrepancy in the dose based on the LAFOV images is driven by the poor quantification of the activity distribution in the PET images.  

Our analysis implies several questions that should be addressed in the future. On one side, a benchmark Simplicit90Y's multi-compartment dose calculation based on LAFOV PET images would be very desirable. A thorough comparison of the accuracy of the post-treatment dosimetry with the pre-treatment dose prediction could indicate a true advantage of LAFOV systems for SIRT.

% \section*{Acknowledgements}%% if any
% Not applicable.

% \section*{Funding}%% if any
% Not applicable.

% \section*{Abbreviations}%% if any
% Text for this section\ldots

% \section*{Availability of data and materials}%% if any
% The datasets used and/or analysed during the current study are available from the corresponding author on reasonable request.

\section*{Ethics approval and consent to participate}%% if any
Not applicable.

\section*{Competing interests}
HS is a full-time employee of Siemens Healthcare AG, Switzerland. AR has received research support and speaker honoraria from Siemens. All other authors have no conflicts of interest to report.

% \section*{Consent for publication}%% if any
% Not applicable.

% \section*{Authors' contributions}
% LM performed the measurements, carried our the analysis and wrote the manuscript. KZ performed the measurements and image reconstructions and revised the manuscript. GAP performed the measurements. HS contributed to the image reconstruction and revised the manuscript. HGR contributed his expertise on Simplicit90Y. AR secured funding. KS conceptualized the study. All authors read and approved the final manuscript. 

%%%%%%%%%%%%%%%%%%%%%%%%%%%%%%%%%%%%%%%%%%%%%%%%%%%%%%%%%%%%%%%%%%%%%%%%%%%%%%%%
%%%%%%%%%%%%%%%%%%%%%%%%%%%%%%%%%%%%%%%%%%%%%%%%%%%%%%%%%%%%%%%%%%%%%%%%%%%%%%%%
\clearpage
\bibliography{sirt_bibliography}

%%%%%%%%%%%%%%%%%%%%%%%%%%%%%%%%%%%%%%%%%%%%%%
% \begin{backmatter}

% \end{backmatter}

\end{document}